# Transport and Magnetism in *p*-type cubic (Ga,Mn)N

KW Edmonds, SV Novikov, M Sawicki[1], RP Campion, C.R. Staddon, AD Giddings, LX Zhao, KY Wang, T Dietl[1], CT Foxon, BL Gallagher

*School of Physics and Astronomy, University of Nottingham, Nottingham NG7 2RD, United Kingdom*
[1]*Institute of Physics, Polish Academy of Sciences, 02-668 Warszawa, Poland*

*The electrical and magnetic properties of p-type cubic (Ga,Mn)N thin films grown by plasma-assisted molecular beam epitaxy are reported. Hole concentrations in excess of $10^{18}$ cm$^{-3}$ at room temperature are observed. Activated behaviour is observed down to around 150K, characterised by an acceptor ionisation energy of around 45-60meV. The dependence of hole concentration and ionisation energy on Mn concentration indicates that the shallow acceptor level is not simply due to unintentional co-doping. Thermopower measurements on freestanding films, CV profilometry, and the dependence of conductivity on thickness and growth temperature, all show that the conduction is not due to diffusion into the substrate. We therefore associate the p-type conductivity with the presence of the Mn in the cubic GaN films. Magnetometry measurements indicate a small room temperature ferromagnetic phase, and a significantly larger magnetic coupling at low temperatures.*

**Introduction**

The emerging field of semiconductor spintronics relies on the ability to manipulate the electron spin in a semiconductor device, thus offering new prospects for non-volatile high speed information storage and processing. An important milestone in this field was the discovery of carrier-mediated ferromagnetism in III-V compounds doped with Mn [1]. Intensive efforts have since led to ferromagnetic transition temperatures $T_C$ in excess of 150K in (Ga,Mn)As [2,3,4], and an impressive range of prototype devices [5,6,7]. However, for widespread technological usage of these systems, a $T_C$ significantly above 300K is necessary, which may yet require the development of new materials.

In this context, the Zener mean-field model prediction of room temperature ferromagnetism in (Ga,Mn)N [8] stimulated much interest. However, an essential ingredient of this model is a *p*-type carrier concentration of around $10^{20}$ cm$^{-3}$, a value significantly higher than the highest hole concentrations so far obtained in GaN. Furthermore, optical measurements have indicated that the Mn acceptor level lies over 1eV above the valence band maximum in Wurtzite (Ga,Mn)N [9,10], which is in agreement with density-of-states calculations [11]. Therefore, in contrast to the case for (Ga,Mn)As, Mn does not appear to be an efficient acceptor in Wurtzite GaN, and may not be expected to interact strongly with delocalised charge carriers in the conduction or valence bands. In spite of this, there are numerous observations of room temperature ferromagnetism in *n*-type (Ga,Mn)N, even in cases where no secondary phases have been identified [12]. The origin of the ferromagnetism in these cases is unresolved, but no conclusive evidence for a coupling between magnetic and semiconductor properties has been demonstrated, and it appears that this effect lies outside the Zener model prediction.

It may be expected that Mn incorporation is favoured in the metastable zincblende (cubic) phase of GaN, since MnN is itself cubic with a similar lattice constant to GaN.

Additionally, cubic GaN does not display the large polar effects associated with Wurtzite material, which may influence incorporation. Even though the Mn level is also predicted to lie deep in the gap in cubic (Ga,Mn)N [11,13], we recently demonstrated that this material can be highly *p*-type, with carrier concentrations exceeding $10^{18}$ cm$^{-3}$ at room temperature [14]. Cubic (Ga,Mn)N may therefore be a more promising candidate than the Wurtzite phase for room temperature carrier-mediated ferromagnetism. It is important to determine the origin of the *p*-type conductivity and investigate the nature of the Mn state in this material. Here we report on a detailed study of the electrical and magnetic properties of cubic (Ga,Mn)N films grown on GaAs(001) by molecular beam epitaxy.

**Growth and structure**

Undoped cubic GaN films and cubic (Ga,Mn)N layers were grown on semi-insulating GaAs (001) substrates by plasma-assisted molecular beam epitaxy (PA-MBE) using arsenic as a surfactant to initiate the growth of cubic phase material [15]. Films where grown under N-rich conditions, which has been found to be necessary for the effective substitutional incorporation of Mn in hexagonal (Ga,Mn)N [16]. The substrate temperature was measured using an optical pyrometer. Growth temperatures from 450 to 680°C were used. The active nitrogen for the growth of the group III-nitrides was provided by an CARS25 RF activated plasma source. The Mn concentration in the films was set using the in-situ beam monitoring ion gauge, and calibrated by secondary ion mass spectrometry (SIMS). The growth was monitored *in-situ* using RHEED.

Prior to the growth of active nitride layers, a GaAs buffer layer was grown on the GaAs substrate in order to provide the cleanest possible interface between GaAs and (Ga,Mn)N, although *p*-type conductivity was also observed in films in which the GaAs buffer was absent. In order to ensure electrical isolation as well as rule out the possibility of Mn diffusion into the GaAs layer being responsible for the observed electrical and magnetic properties, an undoped cubic GaN (~150nm thick) buffer layer followed by a cubic AlN buffer layer (50-150 nm thick) was introduced between the GaAs and cubic (Ga,Mn)N layers in some samples.

The structural properties of the films were studied by x-ray diffraction, using a Philips X'pert Materials Research Diffractometer. 2θ/ω curves for a series of 300nm thick (Ga,Mn)N films grown directly on GaAs buffer layers on GaAs(001) are shown in figure 1. These data confirm that the films are cubic phase and epitaxial with respect to the GaAs substrate, with no peaks corresponding to hexagonal phase GaN visible. Above Mn concentrations of around 5%, an additional peak is visible at around 46.4°, which may correspond to inclusions of GaMn$_3$N or Mn$_4$N secondary phases. Unambiguous identification of the secondary phase is not possible from this measurement, since GaMn$_3$N and Mn$_4$N have nearly the same lattice parameter. Antiferromagnetic GaMn$_3$N precipitates have previously been observed in Wurtzite (Ga,Mn)N at high Mn concentrations [16]. Although Mn$_4$N phases are not usually identified in MBE-grown (Ga,Mn)N, this is a known room temperature ferromagnet, and undetected inclusions of this or other magnetic secondary phases may be responsible for the room temperature ferromagnetic properties observed here as well as elsewhere [17]. For the (Ga,Mn)N (002) reflection, no significant shift in the peak position on varying the Mn concentration between zero and 10% can be resolved, due to the large width of the 002 reflection. (Ga,Mn)N layers grown on cubic AlN buffer layers on GaAs also show only the cubic x-ray diffraction peak, with no obvious degradation of the structural quality resulting from the presence of the buffer layers.

**Transport properties**

Four-point electrical measurements were performed on ~3x3mm squares with evaporated Ti/Al/Ti/Au ohmic contacts in the corners. The contacts were annealed at 440ºC for five minutes. Standard low frequency AC lock-in methods were used for the measurements, with excitation currents typically in the range 100nA-1µA.

Hall effect measurements unambiguously reveal that the cubic (Ga,Mn)N samples are *p*-type. In contrast, nominally undoped GaN films grown under the same N-rich conditions are highly *n*-type, with $n\sim 2 \times 10^{19}$ cm$^{-3}$ at room temperature. Hole density and mobility, obtained from room temperature Hall measurements in fields up to 0.7T, are shown in figure 2 for samples both with and without AlN buffer layers. The hole density $p_{Hall}$ generally increases with increasing Mn up to ~4%, reaching a plateau or decreasing slightly above this value. This plateau coincides with the appearance of a secondary phase in the x-ray diffraction data discussed above. For samples grown with Mn concentration around 0.5 to 1%, denoted by the hashed area in figure 2, we find that samples are highly insulating, and the hole density and mobility cannot be measured accurately. Below 0.5% Mn, samples grown directly on GaAs show a small Hall resistance which changes sign with decreasing temperature, indicating the presence of parallel conducting *n*- and *p*-type regions, while for samples grown on cubic AlN buffer layers, only *p*-type conduction is observed. In the conducting samples, the mobility $\mu_{Hall}$ is consistently in the range 200-350 cm$^2$V$^{-1}$s$^{-1}$, comparable to or larger than the values obtained elsewhere for carbon doped *p*-type cubic GaN [18]. The origin of the observed conducting-insulating-conducting behaviour with increasing Mn concentration is presently unknown, but is reproducible.

The variation of room temperature Hall hole density and mobility in the cubic (Ga,Mn)N layers with growth temperature for constant Mn flux is shown in figure 3. As is evident from figures 2 and 3, the *p*-type behaviour depends on the Mn concentration, but is robust against changes to other growth conditions. Samples grown at temperatures between 450 and 680°C, and also with varying thickness of AlN or GaAs buffer layers, show no systematic variation in the obtained values of hole density and mobility. In addition, we have investigated the effect of varying the film thickness. For thicknesses below around 100nm, the conductivity is significantly decreased, which can be ascribed to the increased influence of interfacial defects. However, for thicker films the conductivity is almost independent of the film thickness, showing that the *p*-type conductivity is a bulk property of the (Ga,Mn)N films rather than being due to surface or interface layers.

Having established the *p*-type conductivity in the cubic (Ga,Mn)N layers, it is important to discuss its origin. Aside from Mn, possible *p*-type dopants in GaN include Mg, Be, and C. Mg or Be contamination can be ruled out as there is no source of Mg or Be in the growth chamber. SIMS measurements identify the presence of unintentional C and O doping in the films, with concentrations in the range $10^{19}$-$10^{20}$cm$^{-3}$. C is a known *p*-type dopant in cubic GaN [18]. However, we find *n*-type conductivity in layers grown without Mn but with similar background C levels. This is a strong indication that the *p*-type conductivity is associated with the presence of Mn.

Since Mn is a well-known *p*-type dopant in GaAs, parallel conduction due to diffusion into the substrate must be considered. The absence of any systematic dependence of the *p*-type conductivity on thickness, growth temperature, or the presence of AlN buffer layers is strong, if not conclusive, evidence against this. To further investigate this possibility, the conduction in the layers close to the surface of the films was investigated by capacitance-voltage profilometry. This yielded a carrier density in the top ≈50nm which is comparable to the value obtained from Hall measurements. However, this technique is rather sensitive to input parameters and film roughness so again cannot be considered conclusive. As a final test, the GaAs substrate was etched away in $H_2O_2$:$H_3PO_4$ solution. Following etching, the films tended to either crack or roll into cylinders due to the very large strain. Four-point Hall measurements could not be performed on such films. However, electrical contacts could be formed at either end of the cylinders using Ag epoxy. The cylinders were found to be electrically conducting, with a room temperature resistance of around $10^5 \Omega$, comparable to the value obtained in unetched films where contacts were made by this method. Seebeck effect measurements were performed on the (Ga,Mn)N cylinders, by placing one end of the cylinders in contact with a heated metal probe. The thermoelectric power was positive for our (Ga,Mn)N cylinders, which demonstrates that the freestanding (Ga,Mn)N layers are *p*-type. Measurements on *p*- and *n*-type GaAs and GaN control samples in the same experimental set-up yielded the expected positive and negative sign of thermoelectric power, respectively. This confirms that the measured *p*-type conductivity is due to the cubic (Ga,Mn)N layer, and not due to diffusion into the GaAs substrate.

Figure 4 shows the temperature-dependence of $p_{Hall}$ and $\mu_{Hall}$ in three cubic (Ga,Mn)N films on AlN, with varying Mn concentration. For all three films, $p_{Hall}$ shows activated behaviour above around 150K, while the mobility increases with decreasing temperature, indicating that phonon scattering is dominant in this regime. The rate of decrease of $p_{Hall}$ is slower than is typically observed in either Mg-doped Wurzite GaN [19,20,21,22] or C-doped zincblende GaN [18], which indicates that the acceptor level is rather shallow in the present samples. We can quantify this using the standard expression:

$$\frac{p(p+N_d)}{(N_a - N_d - p)} = (N_v / g_a)\exp(-\Delta E_a / k_B T) \quad (1)$$

where $E_a$ is the acceptor ionisation energy, $N_a$ and $N_d$ are the acceptor and donor densities, $g_a$ is the acceptor degeneracy which we set equal to 4, and $N_v$ is the effective valence band density of states, given by

$$N_v = 2(2\pi m_h^* k_B T)^{3/2} / h^3 \quad (2)$$

The hole effective mass in GaN, $m_h^*$, is not well known, but is typically found to be in the range $1-2m_o$ [23,24,25]. Here we use an intermediate value of $1.5m_0$, although the choice of $m_h^*$ does not affect the value of $\Delta E_a$ obtained by fitting the experimental data.

From the measured Hall hole densities we obtain values of $\Delta E_a$ of around 45-60meV, which decreases with increasing Mn concentration, as shown in the inset to figure 4. This compares to $\Delta E_a$=215meV [18] for C in cubic GaN. A decreasing $\Delta E$ with increasing acceptor concentration, as found here, is commonly observed in *p*-type GaN [21,22]. $N_a$ and $N_d$ cannot be separately extracted from this fitting procedure, although we obtain the difference ($N_a$-$N_d$) in the range $10^{18}$-$10^{19}$ cm$^{-3}$. This is two orders of magnitude smaller than the Mn concentration measured by SIMS, suggesting that only a fraction of incorporated Mn is electrically active. We expect that compensation is very large in these samples, since nominally undoped samples grown under otherwise identical conditions are strongly *n*-type, with $N_d$~$10^{20}$ cm$^{-3}$. The values obtained from the above fitting procedure should be viewed with caution, since the above equations neglect acceptor level broadening, screening of the acceptors by valence holes, and non-parabolicity of the valence band edge, all of which may be important at these high impurity densities [21,22]. However, the remarkably low value of $\Delta E_a$ obtained, together with its dependence on Mn concentration, provides further evidence that the *p*-type conductivity in these samples is related to the presence of Mn, rather than simply being due to unintentional co-doping by carbon.

At the lowest Mn concentration, freeze-out of carriers occurs rapidly with decreasing temperature, and the conductivity becomes too low to accurately measure below around 220K. At higher concentrations, a change of slope of $p_{Hall}$ is observed at around 150K. Deviations from activated behaviour at low temperatures are frequently observed in *p*-type GaN in which there is significant compensation [19-22], and are usually ascribed to the onset of impurity band conduction [26]. In systems with competing valence and impurity-band conduction, the measured $p_{Hall}$ has a minimum value at the temperature where the conductivity of the valence and impurity band channels is equal, followed by an increase up until $p_{Hall}$ is equal to the impurity band density, which is usually temperature-independent [26]. The measured $p_{Hall}$ can be expressed as a weighted sum of two terms:

$$p_{Hall} = \frac{(\mu_1 p_1 + \mu_2 p_2)^2}{p_1 \mu_1^2 + p_2 \mu_2^2} \qquad (3)$$

where $\mu_1$, $\mu_2$, $p_1$ and $p_2$ are respectively the mobility and density of valence and impurity band carriers.

In the present case, qualitatively different behaviour is observed to those reported in, e.g. refs [19-22]. $p_{Hall}$ reaches a plateau at around 150K, and then at lower temperatures decreases further, although more slowly than observed at higher temperatures. This behaviour can also be reproduced using the expression above, but only if $p_{VB}$ and $p_{IB}$ are both thermally activated (rather than metallic). Figure 5 shows a fit to the data in figure 4 for the sample with 4.2% Mn, with $p_1$ given by equation 1 and $p_2 = p_0 \exp(-\Delta E_2/k_B T)$. We obtain a good agreement with experiment using $\Delta E_a$=50meV, $\Delta E_2$=15meV, $\mu_1/\mu_2$=10, and assuming that the ratio of mobilities is independent of temperature.

The activated carrier density, together with the not too dissimilar mobilities between the two channels, suggests that $p_2$ may not be associated with an impurity band in the present case. The origin of this second conducting channel in these samples is not clear; however it is unlikely to be associated with an interfacial layer, since quantitatively similar behaviour is observed, at this Mn concentration, between samples grown with and without the AlN buffer layer.

**Magnetic properties**

Magnetic measurements are performed in a SQUID magnetometer, with magnetic fields of up to 4000Oe applied in the plane of the sample. Room temperature SQUID measurements for a 4.2% cubic (Ga,Mn)N/AlN/GaAs(001) sample are shown in figure 6. A temperature-independent linear background due to the diamagnetic substrate has been subtracted from the data. Similar to earlier reports of magnetism in (Ga,Mn)N, we find a ferromagnetic signal in all the cubic (Ga,Mn)N films studied here, which persists above 400K. Comparing the measured room temperature ferromagnetism to the Mn concentration measured by SIMS, we obtain a magnetic moment of only around 0.1 $\mu_B$ / Mn, indicating that most of the Mn in the sample is not contributing to the room temperature coupling. The origin of this signal is not presently known, however it has been noted that there are several $Mn_xN_y$ phases which have a ferromagnetic transition temperature above room temperature [17], which could account for the observed behaviour. Here we take the view that, in order to understand the magnetic properties of (Ga,Mn)N, it is important to study the system as a whole rather than concentrating only on the small room temperature ferromagnetic part (we note however that the opposite approach is frequently taken in the experimental literature surrounding (Ga,Mn)N). We therefore study the behaviour at low temperatures, in order to investigate the part of the Mn magnetic moment which does not take part in the high temperature coupling.

The high temperature ferromagnetic phase appears to be temperature-independent below around 200K. Therefore, subtracting magnetisation curves measured at 50K from those measured at lower temperatures should allow us to identify the magnetic behaviour of the remainder of the film. This procedure is followed in figure 7. At low Mn concentration (0.3%), the relative magnetisation $\Delta M(H,T)=M(H,T)-M(H,50K)$ is linear with the external magnetic field both at 15K and 5K, characteristic of paramagnetic behaviour. Different behaviour is observed at 4.2% Mn concentration. A large increase of the magnetisation is observed on going from 15K to 5K, with a clear hysteresis. The effect develops further on lowering of the temperature to 2 K. This is a clear indication of a magnetic coupling between the Mn ions in this sample. Similar magnetisation curves have been reported for insulating hexagonal (Ga,Mn)N films [27], interpreted as having spin-glass characteristics. The glassy behaviour was ascribed to antiferromagnetic superexchange interactions between the substitutional Mn, which are naturally frustrated within the Wurzite lattice. The zincblende lattice also allows for frustrated antiferromagnetism [28], however the Mn concentration in the present case seems too low to expect significant nearest-neighbour interactions.

On the other hand, the present samples exhibit *p*-type conduction, which is expected to favour a *ferromagnetic* alignment of substitutional Mn. Indeed, the magnetisation curves of figure 7 are reminiscent of those observed for highly *p*-type (Zn,Mn)Te single crystals [29]. Even though the samples of ref. [29] exhibited insulating behaviour at low temperatures, recent inelastic neutron measurements indicated local ferromagnetic ordering mediated by weakly localised holes [30]. However, other mechanisms may also give rise to ferromagnetic ordering at low temperatures in *p*-type materials, including percolation of bound magnetic polarons [31], or virtual transitions between valence and impurity bands [32]. Clearly, more work is required to resolve this issue.

**Summary**

We have clearly demonstrated *p*-type conduction in cubic (Ga,Mn)N films grown on GaAs(001) by PA-MBE. Temperature-dependent Hall measurements indicate that the acceptor ionisation energy is around 50meV, which is shallower than any known acceptor level in GaN. The dependence of hole concentration and ionisation energy on the Mn concentration demonstrates than the shallow acceptor level is related to the presence of Mn, although the hole concentration is around two orders of magnitude smaller than the Mn concentration measured by SIMS, probably due to the presence of a large *n*-type background doping. The *p*-type conduction is in contrast to the *n*-type behaviour more usually found in Wurtzite (Ga,Mn)N, and is an essential ingredient for Zener-like carrier-mediated ferromagnetism [8]. Much higher hole concentrations will be necessary to realise the predicted room temperature ferromagnetism according to this model, however the present result suggests that cubic (Ga,Mn)N may be a good candidate system to achieve this. Magnetometry results show the presence of a magnetic ordering at low temperatures (~10K), which is reminiscent of the ferromagnetism mediated by weakly localised holes in (Zn,Mn)Te, and thus may be related to the *p*-type conduction.

**Acknowledgements**
The work was supported by EU projects FENIKS (EC: G5RD-CT-2001-00535) and CELDIS (ICA1-CT-2000-70018), the EPSRC (UK), and Polish KBN grant PBZ-KBN-044/P03/2001. KWE is supported by the Royal Society (UK). We thank Jas Chauhan and Dave Taylor for processing the van der Pauw samples and for etching of the freestanding layers. Valuable discussions with Tomas Jungwirth, Henri Mariette, Joel Cibert, David Ferrand, Piotr Bogusławski, Maria Kaminska, and Andrzej Twardowski are acknowledged.

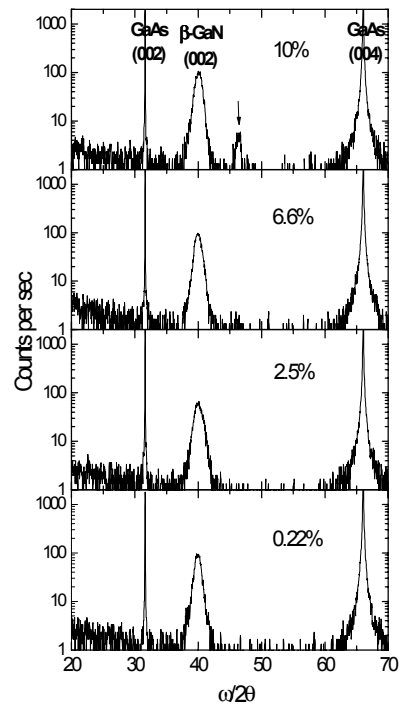

Figure 1. ω-2θ plots of 300nm thick (Ga,Mn)N films on GaAs(001), with Mn concentrations 0.22%, 2.5%, 6.6%, 10%.

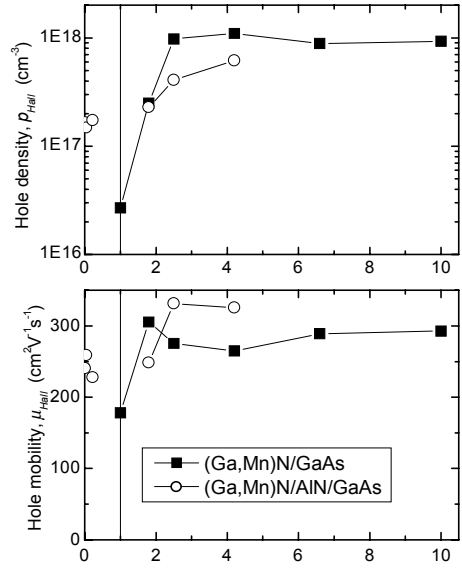

Figure 2. Room temperature Hall hole density and mobility versus Mn concentration for (Ga,Mn)N/GaAs (squares) and (Ga,Mn)N/AlN/GaAs (circles). For both series, samples grown with Mn concentration close to 1%, marked by the hashed region in each figure, are found to have either very low conductivity or be fully insulating at room temperature.

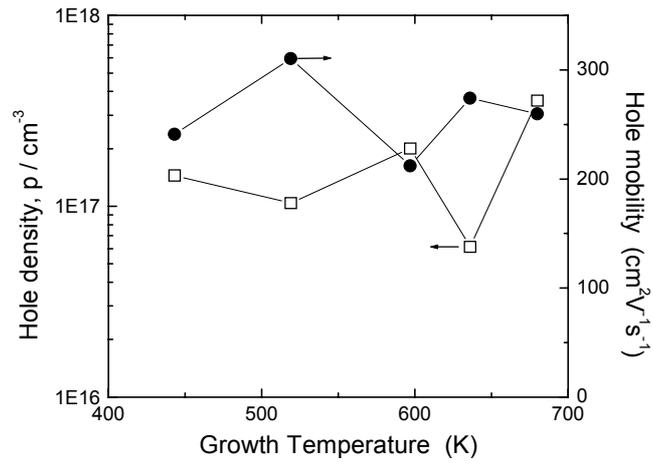

Figure 3. Room temperature Hall hole density and mobility versus growth temperature for a series of cubic (Ga,Mn)N films on GaAs.

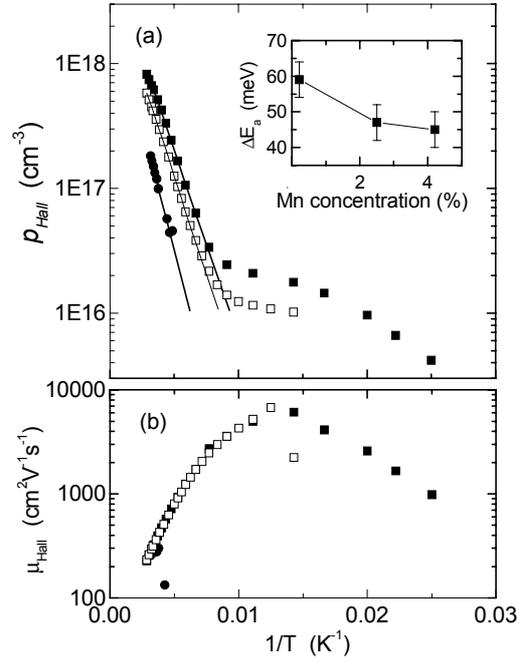

Figure 4. (a) Hole density, and (b) hole mobility, extracted from Hall measurements, versus temperature for cubic (Ga,Mn)N layers on AlN buffers on GaAs(001), for Mn concentrations 4.2% (filled squares), 2.5% (open squares) and 0.22% (circles). The lines in (a) indicate the fitted curves to the activated hole densities, as described in the text. Inset of (a): acceptor ionisation energies extracted from the fits.

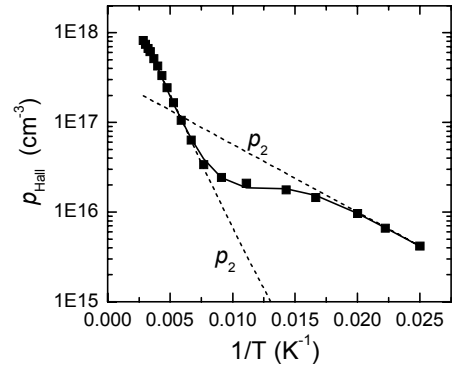

Figure 5. Measured hole density for (Ga,Mn)N/AlN/GaAs(001) (points); calculated $p_{Hall}$ assuming two activated contributions $p_1$ and $p_2$ (lines).

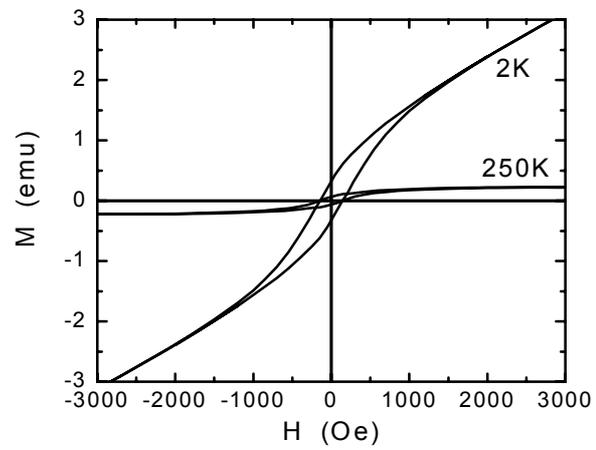

Figure 6. Magnetisation hysteresis curves for a cubic (Ga,Mn)N/AlN/GaAs(001) sample with 4.2% Mn, after substracting a linear background due to the diamagnetic substrate.

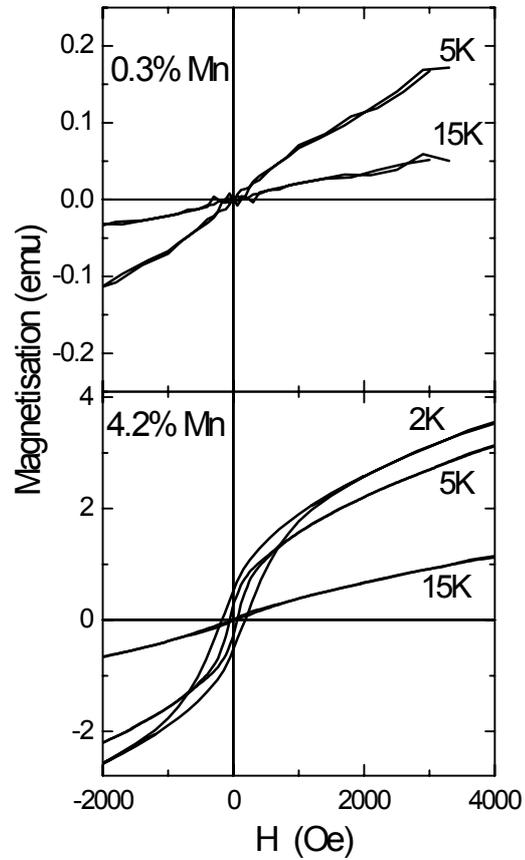

Figure 7. Magnetisation after removing the high $T_C$ contribution, $\Delta M(T)=M(T)-M(50K)$ versus magnetic field, for temperature T=5K and 15K, for (Ga,Mn)N samples with Mn concentration 0.3% (upper panel) and 4.2% (lower panel).